\documentclass[10pt,conference]{IEEEtran}

\begin{document}

\title{Deformed Statistics Formulation of the Information Bottleneck
Method}

\author{
\authorblockN{R. C. Venkatesan}
\authorblockA{Systems Research Corporation\\
Aundh, Pune 411007, India \\
Email: ravi@systemsresearchcorp.com} \and
\authorblockN{A. Plastino}
\authorblockA{IFLP, National University La Plata \\
\& National Research Council (CONICET) \\
C. C., 727 1900, La Plata,Argentina\\
Email: plastino@venus.fisica.unlp.edu.ar} }
 \maketitle

\begin{abstract}
The theoretical basis for a candidate variational principle for the
information bottleneck (IB) method is formulated within the ambit of
the generalized nonadditive statistics of Tsallis. Given a
nonadditivity parameter $ q $, the role of the \textit{additive
duality} of nonadditive statistics ($ q^*=2-q $ ) in relating
Tsallis entropies for ranges of the nonadditivity parameter $ q < 1
$ and $ q > 1 $ is described. Defining $ X $, $ \tilde X $, and $ Y
$ to be the source alphabet, the compressed reproduction alphabet,
and, the \textit{relevance variable} respectively, it is
demonstrated that minimization of a generalized IB (gIB) Lagrangian
defined in terms of the nonadditivity parameter $ q^* $
self-consistently yields the \textit{nonadditive effective
distortion measure} to be the \textit{$ q $-deformed} generalized
Kullback-Leibler divergence: $ D_{K-L}^{q}[p(Y|X)||p(Y|\tilde X)] $.
This result is achieved without enforcing any \textit{a-priori}
assumptions.  Next, it is proven that the $q^*-deformed $
nonadditive free energy of the system is non-negative and convex.
Finally, the update equations for the gIB method are derived. These
results generalize critical features of the IB method to the case of
Tsallis statistics.

\end{abstract}

\section{Introduction}
Rate distortion (RD) theory [1,2] is a major branch of information
theory which provides the theoretical foundations for lossy data
compression.  RD theory addresses the problem of determining the
minimal amount of entropy (or information) $R$ that should be
communicated over a channel, so that the source (input signal/source
alphabet/codebook) $ X\in \mathcal {X} $ can be approximately
reconstructed at the receiver (output signal/reproduction
alphabet/quantized codebook) $\tilde X \in \tilde\mathcal{X} $
without exceeding a given expected distortion $D$. Note that
calligraphic fonts are used to denote sets.  In turn, the
information bottleneck (IB) method is a technique introduced by
Tishby, Pereira, and Bialek [3, 4] for finding the best tradeoff
between accuracy and complexity (compression) when summarizing (e.g.
clustering) a discrete random variable $X$, given a joint
probability distribution between $X$ and a  \textit{relevance
variable} $ Y\in\mathcal{Y} $, i.e. $ p(x,y) $. In this regard, the
IB method represents a significant qualitative improvement over RD
theory.  The IB method has acquired immense utility in machine
learning theory.  For example, the IB method and its modifications
have successfully been employed in applications in diverse areas
such as genome sequence analysis, astrophysics, and, text mining
[4].
 $q-$Deformed (or Tsallis) statistics [5,6] has recently been shown
to yield interesting improvements concerning RD theory [7]. The
present paper extends analogous "q-"considerations to the IB method.

The generalized (nonadditive) statistics of Tsallis' has recently
been the focus of much attention in statistical physics, and allied
disciplines.  Note that the terms generalized statistics.
$q-deformed $ statistics, nonadditive statistics, and nonextensive
statistics are used interchangeably. Nonadditive statistics, which
generalizes the Boltzmann-Gibbs-Shannon (B-G-S) statistics, has
recently found much utility in a wide spectrum of disciplines
ranging from complex systems and condensed matter physics to
financial mathematics. A continually updated bibliography of works
in nonadditive statistics may be found at
http://tsallis.cat.cbpf.br/biblio.htm.

Since the work on nonextensive source coding by Landsberg and Vedral
[8], a number of studies on the information theoretic aspects of
generalized statistics pertinent to coding related problems have
been performed [9-12]. Most recently, the nonadditive statistics of
Tsallis [5,6] has been utilized to develop a generalized statistics
RD theory [7]. This paper [7] investigates nonadditive statistics
within the context of RD theory in lossy data compression.  The
generalized statistics RD model performs variational minimization of
the nonadditive RD Lagrangian employing a method developed [13] to
"rescue" the linear constraints originally employed by Tsallis [5].
Nonadditive statistics possesses a number of constraints having
different forms [14-16].

RD theory is now briefly described as a precursor to introducing the
leitmotif for the IB method.  For a source alphabet $ X \in
\mathcal{X} $ and a reproduction alphabet $ \tilde X \in \tilde
\mathcal{X} $, the mapping of $ x \in \mathcal{X} $ to $ \tilde x
\in \tilde \mathcal{X} $ is characterized by the quantizer $
p(\tilde x|x) $. The RD function is obtained by minimizing the
generalized mutual information (GMI) $ I_{q}(X;\tilde X) $ (defined
in Section 2, [7]) over all normalized $ p\left( {\tilde x\left| x
\right.} \right) $\footnote{The absence of a definitive nonadditive
channel coding theorem sometimes prompts the use of the term
generalized mutual entropy instead of GMI [10].}.   In RD theory $
I_q(X;\tilde X) $ is known as the \textit{compression information}
(see Section 3, [7]). Here, $ q $ is the nonadditivity parameter [5,
6]. A significant feature of the nonadditive RD model [7] is that
the threshold for the compression information is lower than that
encountered in RD models derived from B-G-S statistics. This feature
augurs well for utilizing Tsallis statistics in data compression
applications.

By definition, the nonadditive RD function is [1,2,7]
\begin{equation}
R_q \left( D \right) = \mathop {\min }\limits_{p\left( {\left.
{\tilde x} \right|x} \right):\left\langle {d\left( {x,\tilde x}
\right)} \right\rangle _{p\left( {x,\tilde x} \right)}  \le D} I_q
\left( {X;\tilde X} \right);0<q<1, \
\end{equation}
where, $ R_q(D) $ is the minimum of the \textit{compression
information}.    The \textit{distortion measure} is denoted by $
d(x,\tilde x) $ and is taken to be the Euclidean square distance for
most problems in science and engineering [1,2]. Given $ d(x,\tilde
x) $, the partitioning of $ X $ induced by $ p(\tilde x|x) $ has an
expected distortion $ D=<d(x,\tilde x)>_{p(x,\tilde x)} $. Note that
in this paper, $ \left\langle \bullet \right\rangle_{p(\bullet)} $
denotes the expectation with respect to the probability $ p(\bullet)
$. RD theory \textit{a-priori} specifies the nature of the
distortion measure, which is tantamount to an \textit{a-priori}
specification of the features of interest in the source alphabet $ X
$ to be contained in compressed representation $ \tilde X $.  RD
theory lacks the framework to specify the features of interest in $
X $ to be contained in $ \tilde X $, that are relevant to a given
study. To ameliorate this drawback, the IB method introduces another
variable $ Y \in \mathcal {Y} $, the \textit{relevance variable}.
Note that $ Y $ need not inhabit the same space as $ X $.

\textit{Thus, the crux of the IB method is to simultaneously
minimize the compression information $ I_q(X;\tilde X) $ and
maximize the relevant information $ I_q(\tilde X;Y) $}. More
specifically, the IB method extracts structure from the
\textit{source alphabet} via data compression, followed by a
quantification of the information contained in the extracted
structure with respect to a \textit{relevance variable}.
Consequently, the IB method "squeezes" the information between $ X $
and $ Y $ through a \textit{bottleneck} $ \tilde X $.  The IB method
is compactly described by the Markov condition [3,4]
\begin{equation}
 \tilde X \leftrightarrow X
\leftrightarrow Y .
\end{equation}
As discussed in [7], the un-normalized GMI in Tsallis statistics
acquires different forms in the regimes $ 0 < q < 1 $ and $ q \ge 1
$, respectively. For example, for $ 0 < q < 1 $, the GMI is of the
form $ I_{0<q<1} \left( {X;\tilde X} \right) = -
\sum\limits_{x,\tilde x} {p\left( {x,\tilde x} \right)\ln _q }
\left( {\frac{{p\left( x \right)p\left( {\tilde x}
\right)}}{{p\left( {x,\tilde x} \right)}}} \right) $.

For $ q > 1 $, the GMI is defined by $ I_{q > 1}(X;\tilde
X)=S_q(X)+S_q(\tilde X)-S_q(X,\tilde X) $, where $ S_q(X) $ and $
S_q(\tilde X) $ are the marginal Tsallis entropies for the random
variables $ X $ and $ \tilde X $, and,  $ S_q(X,\tilde X)$ is the
joint Tsallis entropy [7]. Unlike the B-G-S case, $
I_{0<q<1}(X;\tilde X) $ can never acquire the form of $ I_{q
> 1} $, and vice versa [7, 9, 10], the reason being that the
\textit{sub-additivities} $ S_q(X|\tilde X)\leq S_q(X) $ and $
S_q(\tilde X|X)\leq S_q(\tilde X) $ are not generally valid when $
0<q<1 $.  While the form of $ I_{0<q<1}(X;\tilde X) $ is important
in a number of applications of practical interest in coding theory
and learning theory where it is desirable that the GMI be expressed
as the generalized Kullback-Leibler divergence (K-Ld) between the
joint probability $ p(x,\tilde x) $ and the marginal probabilities $
p(x) $ and $ p(\tilde x) $ [1], un-normalized Tsallis entropies for
$ q>1 $ possess a number of important properties such as the
\textit{generalized data processing inequality} and the
\textit{generalized Fano inequality} [10].  The different forms of
the GMI for $ 0<q<1 $ and $ q>1 $ are reconciled by invoking the
\textit{additive duality of nonadditive statistics} [17]. This
entails a re-parameterization of the nonadditivity parameter $ q^* =
2-q $, resulting in \textit{dual Tsallis entropies}.

This paper derives a theoretical basis for the generalized IB (gIB)
method, which is a fundamental qualitative extension of the seminal
work of Tishby, Pereira, and Bialek [3]. This analysis commences
with the minimization of the gIB Lagrangian
\begin{equation}\label{1}
L_{gIB}^q \left[ p({\tilde x\left| x \right.} )\right] \
 = I_{q} \left( {X;\tilde X} \right) - \tilde\beta_{gIB} I_{q} \left( {\tilde X;Y} \right);q>1 ,
\end{equation}
subject to the normalization of $ p\left( {\tilde x\left| x \right.}
\right) $. Here, $ \tilde \beta_{gIB} $ is the \textit{gIB tradeoff
parameter} for the simultaneous minimization and maximization
described by (3). From (3), it is easily shown that: $ \delta
L^q_{gIB}[p(\tilde x|x)]=0 \Rightarrow \frac{{\delta I_q \left(
{X;\tilde X} \right)}}{{\delta I_q \left( {\tilde X;Y} \right)}} =
\frac{1}{{\tilde \beta _{gIB} }} $. Thus, by increasing $
\tilde\beta_{gIB} $, convex curves akin to the RD curves [1,2,7],
may be constructed in the "information plane" ($ I_q(X;\tilde X) $,
$ I_q(\tilde X;Y) $). These are called
\textit{relevance-compression} curves [4].

Apart from its ability to model \textit{long-range interactions}
when performing clustering of complex data sets, the gIB method also
facilitates the analysis of the IB method within the context of
predictability [18]. Predictability may be viewed as an excursion
from the extensive B-G-S statistics, and is inherently nonextensive
(nonadditive).

\section{Tsallis Entropies and Dual Tsallis Entropies}

The un-normalized Tsallis entropy, conditional Tsallis entropy,
joint Tsallis entropy, the jointly convex generalized K-Ld, and, the
GMI
 may thus be written as [19,20]
\begin{equation}
\begin{array}{l}
S_q \left( X \right) =  - \sum\limits_x {p\left( x \right)} ^q \ln
_q p\left( x \right),
 S_q \left( {\left. {\tilde X} \right|X} \right) \\
 =  - \sum\limits_x {\sum\limits_{\tilde x} {p\left( {x,\tilde x} \right)^q \ln _q p\left( {\left. {\tilde x} \right|x} \right)} } , \\
 S_q \left( {X,\tilde X} \right) =  - \sum\limits_x {\sum\limits_{\tilde x} {p\left( {x,\tilde x} \right)^q \ln _q p\left( {x,\tilde x} \right)} } \\
 =S_q(X)+S_q(\tilde X|X)=S_q(\tilde X)+ S_q(X|\tilde X),\\
D_{K-L}^q \left( {p\left( X \right)\left\| {r(X)} \right.} \right) =
- \sum\limits_x {p\left( x \right)}  \ln _q \frac{{r(x)}}{{p(x)}},\\
 I_{0<q<1} \left( {X;\tilde X} \right) =  - \sum\limits_{x,\tilde x} {p\left( {x,\tilde x} \right)\ln _q } \left( {\frac{{p\left( x \right)p\left( {\tilde x} \right)}}{{p\left( {x,\tilde x} \right)}}} \right) \\
 = D_{K - L}^q \left( {p\left( {X,\tilde X} \right)\left\| {p\left( X
\right)p\left( {\tilde X} \right)} \right.} \right),\\
\end{array}
\end{equation}
respectively.  The \textit{q-deformed} logarithm and the
\textit{q-deformed} exponential are defined for $ 0<q<1 $ as [21]
\begin{equation}
\begin{array}{l}
\ln _q \left( x \right) = \frac{{x^{1 - q} - 1}}{{1 - q}}, \\
and, \\
\exp _q \left( x \right) = \left\{ \begin{array}{l}
 \left[ {1 + \left( {1 - q} \right)x} \right]^{\frac{1}{{1 - q}}} ;1 + \left( {1 - q} \right)x \ge 0, \\
 0;otherwise \\
 \end{array} \right.
\end{array}
\end{equation}

The operations of $ q-deformed $ relations are governed by $
q-algebra $ and $ q-calculus $ [21].  Apart from providing an
analogy to equivalent expressions derived from B-G-S statistics, $
q-algebra $ and $ q-calculus $ endow generalized statistics with a
unique information geometric structure. Salient
 results of \textit{q-algebra} employed in this paper involving the $q-deformed  $ addition ($
 \oplus_q $) and subtraction ($ \ominus_q $), are [21]
\begin{equation}
\begin{array}{l}
x\oplus_q y=x+y+(1-q)xy, \\
x\ominus_q y = \frac{{x - y}}{{1 + \left( {1 - q} \right)y}}, where,
\ominus _q y = \frac{{ - y}}{{1 + \left( {1 - q)y} \right)}},\\
 \ln _q \left( {xy} \right) = \ln _q \left( x \right) \oplus _q \ln _q \left( x \right) \\
 = \ln _q \left( x \right) + \ln _q \left( y \right) + \left( {1 - q} \right)\ln _q \left( x \right)\ln _q \left( y \right), \\
 \ln _q \left( {{\raise0.7ex\hbox{$x$} \!\mathord{\left/
 {\vphantom {x y}}\right.\kern-\nulldelimiterspace}
\!\lower0.7ex\hbox{$y$}}} \right) = \ln _q \left( x \right) \ominus _q \ln _q \left( x \right)=y^{q-1}(\ln_q (x)-\ln_q(y)). \\
 \end{array}
\end{equation}
Given two independent variables $ X $ and $ Y $, one of the
fundamental consequences of nonadditivity of the Tsallis entropy is
the \textit{pseudo-additivity} relation
\begin{equation}
S_q \left( {XY} \right) = S_q \left( X \right) + S_q \left( Y
\right) + \left( {1 - q} \right)S_q \left( X \right)S_q \left( Y
\right).
\end{equation}

Re-parameterizing (5) via the \textit{additive duality} $q^*=2-q $,
yields the \textit{dual deformed} logarithm and exponential
\begin{equation}
\begin{array}{l}
 \ln _{q^*}  \left( x \right) =  - \ln _q  \left( {\frac{1}{x}} \right), and, \exp _{q^*}  \left( x \right) = \frac{1}{{\exp _q  \left( { - x} \right)}}. \\
 \end{array}
\end{equation}
A dual Tsallis entropy defined by
\begin{equation}
S_{q^*} \left( X \right) =  - \sum\limits_x {p\left( x \right)\ln
_{q^*} } p\left( x \right).
\end{equation}
The dual Tsallis joint entropy obeys the relation
\begin{equation}
\begin{array}{l}
 S_{q^*} \left( {X,\tilde X} \right) = S_{q^*} \left( X \right) + S_{q^*} \left( {\left. {\tilde X} \right|X} \right), \\
where, \\
S_{q^ *  } \left( {\left. {\tilde X} \right|X} \right) =
-\sum\limits_{x,\tilde x} {p\left( {x,\tilde x} \right)\ln _{q^ *  }
p\left( {\left. {\tilde x} \right|x} \right)}. \\
 \end{array}
\end{equation}
Here, $  \ln_{q^*}(x) = \frac{{x^{1 - q^*} - 1}}{{1 - q^*}} $.
\textit{The dual Tsallis entropies acquire a form identical to the
B-G-S entropies, with $ \ln_{q^*}(\bullet) $ replacing $
\log(\bullet) $}. The GMI's $ I_{q>1}(X;\tilde X) $ and $
I_{0<q^*<1}(X;\tilde X) $ defined by the nonadditivity parameters $
q>1 $ and $ 0<q^*<1 $ respectively, relate to each other as (Theorem
3, [7])
\begin{equation}
\begin{array}{l}
I_{q^*}(X;\tilde X) =  - \sum\limits_x {\sum\limits_{\tilde x} {p\left( {x,\tilde x} \right)\ln _{q^ *  } } } \left( {\frac{{p\left( x \right)p\left( {\tilde x} \right)}}{{p\left( {x,\tilde x} \right)}}} \right) \\
\mathop  = \limits^{\left( q^* \rightarrow q \right)} {S_q \left( X
\right) + S_q \left( {\tilde X} \right) - S_q \left( {X;\tilde X}
\right)} =I_q(X;\tilde X)\\
\mathop  = \limits^{\left( q \rightarrow q^* \right)} I_{q^*}(\tilde X;X). \\
\end{array}
\end{equation}
Here, "$ q^* \rightarrow q  $" is a re-parameterization from $ q^* $
to $ q $, and,"$ q \rightarrow q^* $" is a re-parameterization from
$ q $ to $ q^* $.
\section{Generalized Information Bottleneck Variational Principle}

\subsection{Self-consistent equations}

Depending upon the "upstream" and "downstream" variables in the
Markov condition (2), the total probability may be expressed as
\begin{equation}
\begin{array}{l}
 \tilde X \leftarrow X \leftarrow Y \Rightarrow p\left( {x,\tilde x,y} \right) = p\left( {x,y} \right)p\left( {\left. {\tilde x} \right|x} \right), \\
 \tilde X \to X \to Y \Rightarrow p\left( {x,\tilde x,y} \right) = p\left( {x,\tilde x} \right)p\left( {\left. y \right|x} \right). \\
 \end{array}
\end{equation}
The Markov condition $ \tilde X \leftarrow X \leftarrow Y $ yields
[3]
\begin{equation}
p\left( {\left. y \right|\tilde x} \right) = \frac{1}{{p\left(
{\tilde x} \right)}}\sum\limits_x {p\left( {\left. y \right|x}
\right)p\left( {\left. {\tilde x} \right|x} \right)p\left( x
\right)}.
\end{equation}
Since $ \tilde X \leftarrow X \leftarrow Y = Y \leftarrow X
\leftarrow \tilde X $, the Markov condition yields through
application of Bayes rule and consistency [3]
\begin{equation}
p\left( {\left. y \right|\tilde x} \right) = \sum\limits_{x \in X}
{p\left( {\left. y \right|x} \right)p\left( {\left. {x}
\right|\tilde x} \right)}.
\end{equation}
Thus
\begin{equation}
\begin{array}{l}
 p\left( {\tilde x} \right) = \sum\limits_{x,y} {p\left( {x,\tilde x,y} \right)}  = \sum\limits_x {p\left( x \right)p\left( {\left. \tilde x \right|x} \right),}  \\
 and, \\
 p\left( {\tilde x,y} \right) = \sum\limits_x {p\left( {x,\tilde x,y} \right) = } \sum\limits_x {p\left( {x,y} \right)p\left( {\left. {\tilde x} \right|x} \right)} . \\
 \end{array}
\end{equation}
From (15), the following relations are obtained
\begin{equation}
\frac{{\delta p\left( {\tilde x} \right)}}{{\delta p\left( {\left.
{\tilde x} \right|x} \right)}} = p\left( x \right),and,\frac{{\delta
p\left( {\left. {\tilde x} \right|y} \right)}}{{\delta p\left(
{\left. {\tilde x} \right|x} \right)}} = p\left( {\left. x \right|y}
\right).
\end{equation}
\subsection{The variational principle}
 The gIB Lagrangian (3) cannot be expressed in terms of the generalized K-Ld. As
discussed in [7], the \textit{additive duality} is required to
express the GMI for $ q
> 1 $ in terms of the generalized K-Ld.  This is required
to formulate nonadditive numerical schemes akin to the EM algorithm
[22], using the alternating minimization method based on the
Csisz\'{a}r-Tusn\'{a}dy theory [23]. The gIB Lagrangian in $
q^*-space $ is
\begin{equation}
L_{gIB}^{q^ *  } \left[ {p\left( {\left. {\tilde x} \right|x}
\right)} \right] = I_{q^ *  } \left( {X;\tilde X} \right) - \tilde
\beta_{gIB} I_{q^ *  } \left( {\tilde X;Y} \right);0 < q^ * < 1,
\end{equation}
contingent to the normalization of $ p(\tilde x|x) $. Here, $ I_{q^
*  } \left( {X;\tilde X} \right) $ and $ I_{q^ * } \left( {\tilde
X;Y} \right) $ are obtained from $ I_q(X;\tilde X) $ and $
I_q(\tilde X;Y) $ employing (11). Variational minimization of (17)
[7, 13] yields
\begin{equation}
\begin{array}{l}
 \frac{\delta }{{\delta p\left( {\left. {\tilde x} \right|x} \right)}}L_{gIB}^{q^ *  } \left[ {p\left( {\left. {\tilde x} \right|x} \right)} \right]
 =0 \mathop\Rightarrow \limits^{\left( a \right)} \frac{1}{{q^ *   -
1}}\left( {\frac{{p\left( {\tilde x} \right)}}{{p\left( {\left.
{\tilde x} \right|x} \right)}}} \right)^{1 - q^ *  }  \\
+ \frac{{\tilde \beta _{gIB} }}{{1 - q^ *  }}\sum\limits_y {p\left(
{y\left| x \right.} \right)\left( {\frac{{p\left( y
\right)}}{{p\left( {y\left| {\tilde x} \right.} \right)}}}
\right)^{1 - q^ *  } }  - \frac{{\lambda \left( x \right)}}{{p\left(
x \right)}} = 0.
\end{array}
\end{equation}
Here, $ (a) $ is from Bayes' theorem $ \frac{{p\left( {\left.
{\tilde x} \right|x} \right)}}{{p\left( {\tilde x} \right)}} =
\frac{{p\left( {\left. x \right|\tilde x} \right)}}{{p\left( x
\right)}} $ and $ \frac{{p\left( {\left. {\tilde x} \right|y}
\right)}}{{p\left( {\tilde x} \right)}} = \frac{{p\left( {\left. y
\right|\tilde x} \right)}}{{p\left( y \right)}} $, and, (16). The
term $ p(x) $ is canceled out. A $ \ln_{q^*}(\bullet) $ term is
introduced in (18), by adding and subtracting $ \tilde \beta _{gIB}
\sum\limits_y {\frac{{p\left( {\left. y \right|x} \right)}}{{1 - q^
*  }}} \ $, to yield
\begin{equation}
\begin{array}{l}
\frac{1}{{q^ *   - 1}}\left( {\frac{{p\left( {\tilde x}
\right)}}{{p\left( {\left. {\tilde x} \right|x} \right)}}}
\right)^{1 - q^ *  }  + \tilde \beta_{gIB} \sum\limits_y {p\left(
{y\left| x \right.} \right)\ln _{q^ *  } \left( {\frac{{p\left( y
\right)}}{{p\left( {\left. y \right|\tilde x} \right)}}} \right) } \\
-\lambda^{(1)}=0.
\end{array}
\end{equation}
In (19), $ -\lambda^{(1)}(x)=-\frac{{\lambda \left( x
\right)}}{{p\left( x \right)}} + \tilde \beta _{gIB} \sum\limits_y
{\frac{{p\left( {\left. y \right|x} \right)}}{{1 - q^
*  }}} \ $, which is only dependent on $ x $.
The second term in (19) is expressed as: $ \tilde \beta _{gIB}
\sum\limits_y {p\left( {\left. y \right|x} \right)} \ln _{q^
*  } \left( {\frac{{p\left( {\left. y \right|x} \right)}}{{p\left(
{\left. y \right|\tilde x} \right)}}} \right) \ $ employing $
q^*-deformed $ subtraction and addition (6), yielding
\begin{equation}
\begin{array}{l}
\frac{1}{{q^ *   - 1}}\left( {\frac{{p\left( {\tilde x}
\right)}}{{p\left( {\left. {\tilde x} \right|x} \right)}}}
\right)^{1 - q^ *  } \\
+ \tilde \beta _{gIB}\sum\limits_y {p\left( {\left. y \right|x} \right)} \left[ {\ln _{q^ *  } \left( {\frac{{p\left( y \right)}}{{p\left( {\left. y \right|\tilde x} \right)}}} \right) \ominus_{q^*} \ln _{q^ *  } \left( {\frac{{p\left( y \right)}}{{p\left( {\left. y \right|x} \right)}}} \right)} \right] \\
  - \lambda ^{\left( 1 \right)} \left( x \right) + \tilde \beta _{gIB} \sum\limits_y {p\left( {\left. y \right|x} \right)} \ln _{q^ *  } \left( {\frac{{p\left( y \right)}}{{p\left( {\left. y \right|x} \right)}}} \right)
\oplus_{q^*}=0 \\
 \Rightarrow \frac{1}{{q^ *   - 1}}\left( {\frac{{p\left( {\tilde
x} \right)}}{{p\left( {\left. {\tilde x} \right|x} \right)}}}
\right)^{1 - q^ *  }  + \tilde \beta _{gIB} \sum\limits_y {p\left(
{\left. y \right|x} \right)} \ln _{q^ *  } \left( {\frac{{p\left(
{\left. y \right|x} \right)}}{{p\left( {\left. y \right|\tilde x}
\right)}}} \right) \\
- \lambda ^{\left( 2 \right)} \left( x \right)
= 0. \
\end{array}
\end{equation}
Here, $ - \lambda ^{\left( 2 \right)} \left( x \right) =  -
\left[\tilde \beta _{gIB} {I_{q^ *  } \left( {x;Y} \right)
\oplus_{q^*} \lambda ^{\left( 1 \right)} \left( x \right)} \right] ,
I_{q^ *  } \left( {x;Y} \right) =  - \sum\limits_y {p\left( {\left.
y \right|x} \right)\ln _{q^ *  } \left( {\frac{{p\left( y
\right)}}{{p\left( {\left. y \right|x} \right)}}} \right)} $.
Multiplying (20) by $ p(\tilde x|x) $ and summing over $ \tilde x $,
yields
\begin{equation}
\begin{array}{l}
\lambda ^{\left( 2 \right)} \left( x \right) = \sum\limits_{\tilde
x} {\frac{{p\left( {\left. {\tilde x} \right|x} \right)}}{{q^ *   -
1}}} \left( {\frac{{p\left( {\tilde x} \right)}}{{p\left( {\left.
{\tilde x} \right|x} \right)}}} \right)^{1 - q^ *  } \\
+ \tilde\beta
_{gIB} \left\langle {\sum\limits_y {p\left( {\left. y \right|x}
\right)\ln _{q^ *  } \left( {\frac{{p\left( {\left. y \right|x}
\right)}}{{p\left( {\left. y \right|\tilde x} \right)}}} \right)} }
\right\rangle _{p\left( {\left. {\tilde x} \right|x} \right)} .
\end{array}
\end{equation}

Defining $ \Im _{gIB}(x)  = \left( {q^ *   - 1} \right)\lambda
^{\left( 2 \right)} \left( x \right) $, (20) yields
\begin{equation}
\begin{array}{l}
p\left( {\left. {\tilde x} \right|x} \right) = \\
p\left( {\tilde x}
\right)\frac{{\left\{ {\left[ {1 - \left( {q^ *   - 1}
\right)\frac{{\tilde \beta_{gIB} }}{{\Im _{gIB} \left( x
\right)}}\sum\limits_y {p\left( {y\left| x \right.} \right)\ln _{q^
*  } \left( {\frac{{p\left( {\left. y \right|x} \right)}}{{p\left(
{\left. y \right|\tilde x} \right)}}} \right)} } \right]}
\right\}^{\frac{1}{{^{q^ *   - 1} }}} }}{{\Im _{gIB} \left( x
\right)^{\frac{1}{{1 - q^ *  }}} }}.
\end{array}
\end{equation}
Setting $ {\frac{{\tilde \beta_{gIB} }}{{\Im _{gIB} \left( x
\right)}}} = \tilde\beta_{gIB}(x) $, and, invoking the
\textit{additive duality} in the numerator ($ q^{*}=(2-q) $), (22)
yields the canonical transition probability
\begin{equation}
\begin{array}{l}
 p\left( {\left. {\tilde x} \right|x} \right) = p\left( {\tilde x} \right)\frac{{\exp _q \left[ { - \tilde \beta _{gIB} \left( x \right)D_{K -
L}^q \left[ {p\left( {\left. y \right|x} \right)\left\| {p\left(
{\left. y \right|\tilde x} \right)} \right.} \right]}
\right]}}{{\tilde Z\left( {x,\tilde \beta _{gIB} \left( x \right)}
 \right)}}\\
 \Rightarrow \tilde Z\left( {x,\tilde \beta _{gIB} \left( x \right)}
\right) = \Im _{gIB} \left( x \right)^{\frac{1}{{1 - q^ *  }}}.
\end{array}
\end{equation}
In (23), \textit{$ \tilde \beta _{gIB}(x) $ is the gIB tradeoff
parameter evaluated for each source alphabet $ x \in X $}, and, $
{\tilde Z\left( {x,\tilde \beta _{gIB} \left( x \right)} \right)} $
is the partition function. \textit{The effective distortion measure
has been self consistently obtained via the variational principle to
be $ {D_{K - L}^q \left[ {\left. {p\left( {\left. y \right|x}
\right)} \right\|p\left( {\left. y \right|\tilde x} \right)}
\right]} $, without any a-priori assumptions.} In the limit $ q
\rightarrow 1 $ the B-G-S statistics result [3,4] is recovered.
Solutions of (23) are valid only for $ \left\{ {1 - \left( {1 - q}
\right) \tilde \beta _{gIB}(x)D_{K - L}^q \left[ {\left. {p\left(
{\left. y \right|x} \right)} \right\|p\left( {\left. y \right|\tilde
x} \right)} \right]} \right\} > 0 $.  The condition $ \left\{ {1 -
\left( {1 - q} \right) \tilde \beta _{gIB}(x)D_{K - L}^q \left[
{\left. {p\left( {\left. y \right|x} \right)} \right\|p\left(
{\left. y \right|\tilde x} \right)} \right]} \right\} < 0 $ is
called the \textit{Tsallis cut-off condition} [5], and requires
setting $ p(\tilde x|x) $ =0 and stopping the iteration at the given
$ \tilde\beta_{gIB} $.

\subsection{Free energy of the system}
The $ q^{*}-deformed $ \textit{nonadditive free energy of the
system} is [3]
\begin{equation}
\begin{array}{l}
F_{gIB}^{q^ *  }[p(\tilde x|x);p(\tilde x);p(y|\tilde x)] =F_{gIB}^{q^ *  }[\bullet]\\
= -\left\langle {\ln _{q^ *  } \tilde Z\left( {x,\tilde \beta _{gIB}
\left( x \right)} \right)} \right\rangle _{p\left( x \right)}=
\left\langle {\frac{{\Im _{gIB} \left( x \right) - 1}}{{q^
*   - 1}}} \right\rangle _{p\left( x \right)}.
\end{array}
\end{equation}
Here, $ \left[ \bullet  \right] $ denotes the arguments $ {\left[
{p\left( {\left. {\tilde x} \right|x} \right);p\left( {\tilde x}
\right);p\left( {\left. y \right|\tilde x} \right)} \right]} $ of
the free energy.  Note $ F_{gIB}^{q^ *  }[\bullet] = \tilde \beta
_{gIB}F^{Helmholtz}_{q^*} $, where $ F^{Helmholtz}_{q^*} $ is the $
q^*-deformed $ gIB Helmholtz free energy. Invoking (4) and (21),
(24) yields
\begin{equation}
\begin{array}{l}
 F_{gIB}^{q^ *  }[\bullet] = I_{q^ *  } \left( {X;\tilde X}
 \right)+ \\
\tilde \beta _{gIB} \left\langle {\sum\limits_y {p\left( {\left. y
\right|x} \right)\ln _{q^ *  } \left( {\frac{{p\left( {\left. y
\right|x} \right)}}{{p\left( {\left. y \right|\tilde x} \right)}}}
\right)} } \right\rangle _{p\left( {x,\tilde x}
 \right)} \\
\mathop  = \limits^{\left( a \right)} I_{q^ *  } \left( {X;\tilde X} \right) - \tilde \beta _{gIB}\left\langle { \sum\limits_y {p\left( {\left. y \right|x} \right)} \ln _q \left( {\frac{{p\left( {\left. y \right|\tilde x} \right)}}{{p\left( {\left. y \right|x} \right)}}} \right)} \right\rangle _{p\left( {x,\tilde x} \right)}  \\
\mathop  = \limits^{\left( b \right)} I_{q^ *  } \left( {X;\tilde X} \right) - \tilde \beta _{gIB} \sum\limits_{x,\tilde x,y} {p\left( {x,\tilde x,y} \right)} \ln _q \left( {\frac{{p\left( {x,\tilde x} \right)p\left( y|\tilde x \right)}}{{p\left( {x,\tilde x,y} \right)}}} \right) \\
= D_{K - L}^{q^* } \left[ {p\left( {X,\tilde X} \right)\left\|
{p\left( X \right)p\left( {\tilde X} \right)} \right.}
  \right] \\
  + \tilde \beta _{gIB}D_{K - L}^q \left[ {p\left( {X,\tilde X,Y} \right)\left\| {p\left( {X,\tilde X} \right)p\left( Y|\tilde X \right)} \right.} \right]. \\
 \end{array}
\end{equation}
Here, $ (a) $ invokes the \textit{additive duality }in the second
term in order to introduce the \textit{expected effective
distortion} in $ q-space $, and, $ (b) $ invokes (12) to obtain the
total probability $ p(x,\tilde x,y) $. In (25), $
F_{gIB}^{q^*}[\bullet] $ is the sum of two generalized K-Ld's having
nonadditivity parameters $ q^* $ and $ q $, where $ 0<q^*<1$ and $
q>1 $.  From [24], it is readily follows that $
F_{gIB}^{q^*}[\bullet] $ is non-negative and convex. The expected
effective distortion term in (25) is related to the relevant
information as
\begin{equation}
\begin{array}{l}
 F_{gIB}^{q^*} [\bullet]  = I_{q^ *  } \left( {X;\tilde X} \right) \\
- \tilde \beta _{gIB} \underbrace {\sum\limits_{x,\tilde x} {p\left(
{x,\tilde x} \right)\sum\limits_y {p\left( {\left. y \right|x}
\right)} } \ln _q \left( {\frac{{p\left( {\left. y \right|\tilde x}
\right)}}{{p\left( {\left. y \right|x} \right)}}} \right)}_{D_{eff}
} \\
 \mathop  = \limits^{\left( a \right)} I_{q^ *  } \left( {X;\tilde X} \right) \\
  - \tilde \beta _{gIB} \sum\limits_{x,\tilde x,y} {p\left( {x,\tilde x,y} \right)^q \left[ {\ln _q p(y|\tilde x) - \ln _q p(y|x)} \right]}  \\
 \mathop  = \limits^{\left( b \right)} I_{q^ *  } \left( {X;\tilde X} \right) + \tilde \beta _{gIB} \left[ {I_q \left( {X;Y} \right) - I_q \left( {\tilde X;Y} \right)} \right] \\
  \Rightarrow D_{eff}
=I_q \left( {X;Y} \right) - I_q \left( {\tilde X;Y} \right). \\
 \end{array}
\end{equation}
Note $ I_q \left( {\tilde X;Y} \right) \le I_q \left( {X;Y} \right)
$ by the \textit{generalized data processing inequality} [10]. Here,
$ (a) $ employs $ \ln_q(x/y)=y^{q-1}(\ln_q(x)-\ln_q(y)) $ in (6),
$(b) $ adds and subtracts $ S_q(Y) $, and, invokes (11) and the
symmetry of the GMI: $ I_q(Y;X)=I_q(X;Y);I_q(Y;\tilde X)=I_q(\tilde
X;Y) $. \textit{From (22)-(24), an empirical criterion equivalent to
the Tsallis cut-off condition, described in terms of the gIB free
energy for any $ x\in X $, is: $ 1 + \left( {q^*   - 1}
\right)F_{gIB}^{q^* } \left[ \bullet \right]\left( x \right)<0 $}.
\section{The Update Equations}
\textbf{Lemma 1}:  Given a joint distribution $ p(x)p(\tilde x|x) $,
the distribution $ p(\tilde x) $ that minimizes $
D^q_{K-L}[p(X)p(\tilde X|X)||p(X)p(\tilde X)] $ is the marginal $ p^
*  \left( {\tilde x} \right) = \sum\limits_x {p(x)p(\tilde x|x)} $,
i.e.
\begin{equation}
\begin{array}{l}
 D_{K - L}^{q^*} \left[ {p\left( X \right)p\left( {\left. {\tilde X} \right|X} \right)\left\| {p\left( X \right)p^ *  \left( {\tilde X} \right)} \right.} \right] \\
= \mathop {\min }\limits_{p\left( {\tilde x} \right)} D_{K - L}^{q^*} \left[ {p\left( X \right)p\left( {\left. {\tilde X} \right|X} \right)\left\| {p\left( X \right)p\left( {\tilde X} \right)} \right.} \right]. \\
 \end{array}
\end{equation}
Also
\begin{equation}
\begin{array}{l}
\left\langle {D_{K - L}^q \left[ {\left. {p\left( {\left. {\tilde x}
\right|x} \right)} \right\|p^ *  \left( {\tilde x} \right)} \right]}
\right\rangle _{p\left( x \right)} = \\
\mathop {\min
}\limits_{p\left( {\tilde x} \right)} \left\langle {D_{K - L}^q
\left[ {\left. {p\left( {\left. {\tilde x} \right|x} \right)}
\right\|p\left( {\tilde x} \right)} \right]} \right\rangle _{p\left(
x \right)}.
\end{array}
\end{equation}
\textbf{Proof:}  The positivity condition for (27) is proven in
Lemma 1 of [7], with $ q^* $ replacing $ q $. The positivity
condition for (28) is
\begin{equation}
\begin{array}{l}
 -\sum\limits_{x,\tilde x} {p\left( {x,\tilde x} \right)\ln _q } \left( {\frac{{p\left( {\tilde x} \right)}}{{p\left( {\left. {\tilde x} \right|x} \right)}}} \right) + \sum\limits_{x,\tilde x} {p\left( {x,\tilde x} \right)\ln _q } \left( {\frac{{p^ *  \left( {\tilde x} \right)}}{{p\left( {\left. {\tilde x} \right|x} \right)}}} \right) \\
\mathop  = \limits^{\left( a \right)}  - \sum\limits_{x,\tilde x} {p\left( {x,\tilde x} \right)} \ln _{q^ *  } \left( {\frac{{p\left( {\left. {\tilde x} \right|x} \right)}}{{p^ *  \left( {\tilde x} \right)}}} \right) + \sum\limits_{x,\tilde x} {p\left( {x,\tilde x} \right)} \ln _{q^ *  } \left( {\frac{{p\left( {\left. {\tilde x} \right|x} \right)}}{{p\left( {\tilde x} \right)}}} \right) \\
\mathop  = \limits^{\left( b \right)} D_{K - L}^{q^ *  } \left[
{\left. {p^ *  \left( {\tilde x} \right)} \right\|p\left( {\tilde x}
\right)} \right] \\
  \times \left[ {1 + \left( {q - 1} \right)\left\langle {D_{K - L}^q \left[ {\left. {p\left( {\left. {\tilde x} \right|x} \right)} \right\|p\left( {\tilde x} \right)} \right]} \right\rangle _{p\left( x \right)} } \right]>0;\\
  \forall 0<q^*<1,q>1. \\
 \end{array}
\end{equation}
Here, $ (a) $ invokes the \textit{additive duality}, $ (b) $ employs
the \textit{q-deformed} algebra definition for $ \ln _{q^*} \left(
{\frac{a}{b}} \right) $ from (6) [21] by multiplying and dividing by
$ \frac{{\left[ {\sum\limits_{x,\tilde x} {p\left( {x,\tilde x}
\right)}  + \left( {1 - q^ *  } \right)\sum\limits_{x,\tilde x}
{p\left( {x,\tilde x} \right)} \ln _{q^ *  } \left( {\frac{{p\left(
{\left. {\tilde x} \right|x} \right)}}{{p\left( {\tilde x}
\right)}}} \right)} \right]}}{{\sum\limits_{x,\tilde x} {p\left(
{x,\tilde x} \right)} }}  $, $ {\sum\limits_{x,\tilde x} {p\left(
{x,\tilde x} \right)} } =1 $, and, establishes $ p^
* \left( {\tilde x} \right) = \sum\limits_x {p\left( x
\right)p\left( {\left. {\tilde x} \right|x} \right)} $ after
subjecting the term within brackets ($[\bullet]$) in (29) to the
\textit{additive duality} ($q=2-q^* $).

 The free energy $ F_{gIB}^{q^* }[\bullet] $ is convex only when independently evaluated with
respect to one of the convex distribution sets  $ \left\{ {p\left(
{\left. {\tilde x} \right|x} \right)} \right\} $, $ \left\{ {p\left(
{\tilde x} \right)} \right\} $, and $ \left\{ {p\left( {\left. y
\right|\tilde x} \right)} \right\} $. \textit{The update equations
which minimize the free energy are obtained by projecting the free
energy onto each convex distribution while keeping the other two
arguments constant}.

\textbf{Theorem 1:}  Equations (14), (15) and (23) are satisfied at
the minima of the free energy (24) for each argument of the free
energy as
\begin{equation}
\mathop {\min }\limits_{p\left( {\left. y \right|\tilde x} \right)}
\mathop {\min }\limits_{p\left( {\tilde x} \right)} \mathop {\min
}\limits_{p\left( {\left. {\tilde x} \right|x} \right)} F_{gIB}^{q^
* } \left[ {p\left( {\left. {\tilde x} \right|x} \right);p\left(
{\tilde x} \right);p\left( {\left. y \right|\tilde x} \right)}
\right].
\end{equation}
Denoting the iteration level as $ (\tau) $, minimization is
performed independently by converging alternating iterations
\begin{equation}
\begin{array}{l}
p^{\left( {\tau+1 } \right)} \left( {\left. {\tilde x} \right|x}
\right) \leftarrow p^{\left( \tau  \right)} \left( {\tilde x}
\right)\frac{{\exp _q \left[ { - \tilde \beta _{gIB}^{(\tau)} \left(
x \right)D_{K - L}^q \left[ {\left. {p\left( {\left. y \right|x}
\right)} \right\|p^{\left( \tau  \right)} \left( {\left. y
\right|\tilde x} \right)} \right]} \right]}}{{\tilde Z^{\left(
{\tau+1} \right)} \left( {x,\tilde \beta _{gIB} \left( x \right)}
\right)}},\\
where, \tilde \beta _{gIB}^{\left( \tau  \right)} \left( x \right) =
\frac{{\tilde \beta _{gIB} }}{{\Im _{gIB}^{\left( \tau \right)}
\left( x \right)}} = \frac{{\tilde \beta _{gIB} }}{{\tilde Z^{\left(
\tau  \right)} \left( {x,\tilde \beta _{gIB} \left( x
\right)}\right)^{1 - q^ *  } }}, \\
p^{\left( {\tau  + 1} \right)} \left( {\tilde x} \right) \leftarrow \sum\limits_x {p\left( x \right)p^{\left( \tau+1  \right)} \left( {\left. {\tilde x} \right|x} \right)} , \\
p^{\left( {\tau  + 1} \right)} \left( {\left. y \right|\tilde x}
\right) \leftarrow \frac{1}{{p^{\left( {\tau  + 1} \right)} \left(
{\tilde x} \right)}}\sum\limits_x {p\left( {x,y} \right)p^{\left(
{\tau  + 1} \right)} \left( {\left. {\tilde x} \right|x} \right)}. \\
\end{array}
\end{equation}

\textbf{Proof:} The outline of the proof is presented herein owing
to space constraints.  Defining $ \tilde F_1^{q^
*  } \left[ \bullet  \right] = F^{q^ *  }_{gIB} \left[ \bullet  \right] +
\tilde\lambda \left( x \right)\left( {p\left( {\left. {\tilde x}
\right|x} \right) - 1} \right) $, and following the procedure in
Section III.B. of this paper, $ \frac{{\delta \tilde F_1^{{q^
*  } } \left[  \bullet  \right]}}{{\delta p\left( {\left. {\tilde x}
\right|x} \right)}} = 0 $ exactly yields (26). Minimization with
respect to $ p(y|\tilde x) $ affects only $ D_{eff} $ in (25).
Defining $ \tilde F_2^{q^
* } \left[ \bullet \right] = F^{q^
* }_{gIB} \left[ \bullet  \right] + \tilde \lambda \left( {\tilde x}
\right)\left( {p\left( {\left. y \right|\tilde x} \right) - 1}
\right) $ and invoking $ D_{eff}=I_q(X;Y)-I_q(\tilde X;Y) $ from
(26), employing (4), (11), (12), and (15) yields
\begin{equation}
\begin{array}{l}
  - \tilde \beta _{gIB} \sum\limits_x {p\left( {x,\tilde x,y} \right)^q } \frac{{\delta \ln _q p\left( {\left. y \right|\tilde x} \right)}}{{\delta p\left( {\left. y \right|\tilde x} \right)}} + \tilde \lambda \left( {\tilde x} \right) = 0 \\
  \Rightarrow  - \frac{{\frac{1}{{p\left( {\tilde x} \right)^q }}\sum\limits_x {p\left( {x,y} \right)^q p\left( {\left. {\tilde x} \right|x} \right)^q } }}{{p\left( {\left. y \right|\tilde x} \right)^q }} + \tilde \lambda _1 \left( {\tilde x} \right) = 0;\tilde\lambda_1={  \frac{{\tilde \lambda \left( {\tilde x} \right)}}{{\tilde \beta _{gIB} }}}  \\
  \Rightarrow p\left( {\left. y \right|\tilde x} \right) = \frac{1}{{p\left( {\tilde x} \right)}}\sum\limits_x {p\left( {x,y} \right)p\left( {\left. {\tilde x} \right|x} \right)}.  \\
 \end{array}
\end{equation}

From (27), it may be shown that $ p(\tilde x) $ minimizes $
I_{q^*}(X;\tilde X) $.  Since, $ D_{eff} $ is the expectation of a
generalized K-Ld, (28) is applied to demonstrate that $ p(\tilde x)
$ is a minimizer of $ D_{eff} $.  Note that the gIB update equations
are not globally covergent.
\section{Conclusions and Discussions}
Akin to the RD theory, the degree of compression may be assessed by
the \textit{compression information} $ I_{q^*}(X;\tilde X) $.
However, while the RD method is \textit{upper bounded} by an
\textit{a-priori} chosen optimal expected distortion $ D $, the gIB
method is \textit{lower bounded} by the \textit{relevant
information} $ I_{q^*}(\tilde X;Y) $.  It has been demonstrated that
in lossy compression, $ I_{q^*}(X;\tilde X) $ is always lower than
its counterpart obtained using B-G-S statistics [7]. This
observation implies that gIB \textit{relevance-compression} curves
will tend to traverse the \textit{forbidden region} of an equivalent
IB method based on B-G-S statistics.  Future work casts the gIB
model within the framework of Bregman divergences [25].

\section*{Acknowledgment}
RCV gratefully acknowledges support from \textit{RAND-MSR} contract
\textit{CSM-DI $ \ \& $ S-QIT-101155-03-2009}.



%

\end{document}